\documentclass[letterpaper, 12 pt, conference]{ieeeconf} 
\usepackage[utf8]{inputenc}
\usepackage[noadjust]{cite}
\usepackage{algorithm}
\usepackage[noend]{algpseudocode}
\usepackage{comment}
\usepackage{graphicx}
\graphicspath{ {./images} }
\usepackage{subcaption}
\usepackage{amsmath}
\usepackage{amssymb}
\usepackage{color}
\usepackage{mathabx}
\usepackage{array}
\usepackage{verbatim}
\usepackage{authblk}
\title{Meeting Users’ QoS in a Edge-to-Cloud Platform via Optimally Placing Services and Scheduling Tasks
}
\author[1]{Matthew Turner}
\author[1]{Hana Khamfroush}
\affil[1]{Department of Computer Science, University of Kentucky}
\begin{document}
\maketitle
\begin{abstract}
This paper considers the problem of service placement and task scheduling on a three-tiered edge-to-cloud platform when user requests must be met by a certain deadline. Time-sensitive applications (e.g., augmented reality, gaming, real-time video analysis) have tight constraints that must be met. With multiple possible computation centers, the ``where" and ``when" of solving these requests becomes paramount when meeting their deadlines. We formulate the problem of meeting users' deadlines while minimizing the total cost of the edge-to-cloud service provider as an Integer Linear Programming (ILP) problem. We show the NP-hardness of this problem, and propose two heuristics based on making decisions on a local vs global scale. We vary the number of users, the QoS constraint, and the cost difference between remote cloud and cloudlets(edge clouds), and run multiple Monte-Carlo runs for each case. Our simulation results show that the proposed heuristics are performing close to optimal while reducing the complexity.
\end{abstract}


\captionsetup[figure]{font=small,skip=0pt}

\section{Introduction}
Recent years have seen the focus of computing shift to a centralized approach utilizing cloud technology. While the cloud offers numerous advantages for users, there are always limitations in a centralized system, including security and latency issues. Furthermore, many Internet-of-Things (IoT) applications require a very short response time that might not be possible with conventional centralized data centers. An emerging focus of interest in both industry and academia is finding the optimal way to move the dematerialized strengths of the cloud to a decentralized architecture~\cite{ons-vision}.

The utilization of finite edge devices has become the prevailing solution to the limitations of the centralized cloud. Edge devices here are considered smaller computational devices which are geographically closer to a user-base than the cloud~\cite{ons-service}. These devices are used in cases when a user's device lacks the power to solve a task and/or the remote cloud cannot serve the request in a time-sensitive manner~\cite{ons-migration}. We consider a three-layer system, the "\textit{edge-to-cloud platform}", as the combination of layers including: the users, the cloudlets (edge devices), and the remote cloud ~\cite{khamfroush1}. For simplicity's sake, we will refer to the cloud and cloudlets as "computational devices". Each user has a request that must be served by a computational device, and has a certain amount of time they are willing to wait for that request. Which is the best computational device to serve each task and host each needed service? This paper aims to outline the problem of placing services and scheduling user requests, as well as provide novel heuristics to approach this problem.

To the best of our knowledge, our paper is the first to consider a joint service placement and request scheduling problem given explicitly defined quality-of-service (QoS) and hardware constraints in a three-tiered edge-to-cloud platform with the intention of meeting all users' QoS while minimizing cost to the service provider.
\section{Related Work}
Service placement is a problem with significant interest in the current literature. In~\cite{ons-multiuser}, authors considered the energy cost incurred by offloading tasks from the user onto a cloudlet. In particular, this work proposed a sub-optimal algorithm for approaching when offloading should occur (based on the energy cost), but does not consider the users' QoS constraints. Another approach to offloading is using a probabilistic model based on Markov chains~\cite{ons-followme}. The offloading decision factors in the user's geodesic location in relation to base stations which serve as cloudlets for their example. Over the other model, this work offered the advantage of considering offloading between cloudlets as well. However, this model did not consider the users' QoS constraints, similar to the other work. For both of these works, this implies there is no guarantee for runtime of user tasks.

As well as placement, service scheduling and migration are another area of particular interest. Considering quality of service,~\cite{ons-service} dealt with unpacking and scheduling mobile applications in a ``edge-to-user" system. The work also proposed an algorithm for scheduling the tasks on one of a number of cloudlets. However, they do not consider the heterogeneous limitations of the cloudlets, leading to an unrealistic scenario for the service provider. Serving mobile users require offloading and introduce unique service allocation problems~\cite{ons-migration}. Despite their consideration of user mobility, they do not consider the cost incurred to the service provider for offering these services. Adaptation of this model by service providers seems slim for this reason, since their interests aren't addressed.

As well as migration, static task distribution is a large area of interest. Sufficient work has been done into optimal task distribution for edge-cloud optimization~\cite{ons-distrib}~\cite{khamfroush1}~\cite{khamfroush2}.
In particular,~\cite{ons-distrib} pertains to satisfying user QoS, while increasing the number of tasks that can be handled by the edge-cloud infrastructure. However, this work does not consider the energy consumption or the costs to the service provider and cloudlets.

Some other work has considered both service placement, and request scheduling. For instance, ~\cite{ons-fog} considers the QoS constraints the user, realistic cloud and service constrains, and the problem of scheduling user tasks. This is achieved by utilizing an Integer Linear Programming (ILP) formulation and optimizing it. However, this work does not consider the problem or potential benefits of service placement prior to request scheduling. In reality, most user requests are one of a finite number of services, and the potential computation saving by not redeploying matters to the service provider.  Similarly, ~\cite{ons-fogplan} has similar considerations. Again, an optimization problem considering both service placement, and task selection is crafted and then approximated through a novel algorithm. This work even goes as far to consider redeployment and sharing of services, as well as optimizing service provider cost. However, their proposed algorithms do not consider both the user cost and user QoS, opting to have an algorithm for each interest.

In our previous work, we had considered the problem of joint service placement and task scheduling on a multi-tiered network~\cite{khamfroush1}~\cite{khamfroush2}. In both works, authors proposed a joint placement and scheduling algorithm under realistic hardware constraints. In~\cite{khamfroush1}, the problem was introduced and proven to be NP-Hard; then, an algorithm to approximate the optimal placement and scheduling was provided which ran only under certain conditions. This work was improved upon in~\cite{khamfroush2}, where a more general algorithm was developed to place and schedule while maximizing the number of users served. This work did not examine cost to the service provider, and it also did not meet all users' QoS constraints.

As cloudlets and the remote cloud may both being owned by the same service provider or leased from other service providers, understanding the tradeoff between the cost of service provider and the QoS for the user is becoming a challenge. To this end, we consider a joint optimization problem addressing the trade-off by jointly optimizing service placement on the edge clouds (cloudlets), and scheduling users' requests.

\section{Proposed Contribution}
Much of the current body of work above seeks to minimize energy cost to the service provider, or maximize the users' QoS. Our objective is to combine the two approaches to achieve a balance in the trade-off of users' QoS and service provider cost. Here, cost to the service provider is expressed as a function of the \textit{number of services} that they \textit{need to place} on the \textit{computational devices}, as well as the cost of serving the users' requests on a particular device. QoS is expressed as a function of the time that users must wait for their tasks to be completed. We assume that performing a task on the cloudlet is inherently better for the user than performing it on the cloud, due to the lower latency of communication. However, it adds to the cost of service providers. Thus, in our service placement, we seek to find the number of tasks that should be completed on one of the finite cloudlets while maintaining that all users' QoS constraints are met. We also work under the reality that cloudlets have limited storage and computational abilities. As the user should not see a difference in service due to \textit{where} their task is being scheduled, we assume that they connect to one cloudlet: their \textit{local} cloudlet. We factor in the time of communication between a users' local cloudlet and the computational device that serves their request. We see this as a unique viewpoint and natural combination of the works that have been developed in the field thus far.

\section{Problem Formulation}
\subsection{Network model}
\begin{figure}
  \centering
  \includegraphics[scale=.4]{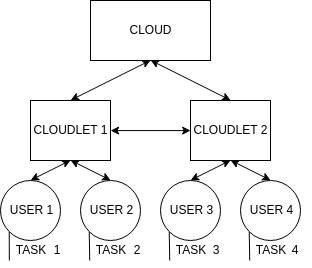}
  \vspace{0.25cm}
  \caption{Example of ``edge-to-cloud platform"}
  \label{fig:my_label}
\end{figure}
\setlength{\belowcaptionskip}{-10pt}
We consider the \textit{edge-to-cloud platform}. Each user has a single task to be completed. We assume that each task fits one of a number of service types, and that it arrives to a computational device all at once. Each user is explicitly connected to a single cloudlet, and they are not aware of where their request is actually solved. For a task to be completed on a computing device, the corresponding service must be on that device. We assume that putting a service on a computational device takes some storage on that device, which limits us from placing an infinite number of services on the cloudlets. The cloud however, has a comparably infinite storage and processing power, therefore, the cloud can always has the services requested by the users. Each user has an associated quality of service threshold, which translates to the total amount of time they're willing to wait for their request to be completed. When they send their request to the cloudlet, the cloudlet serves the request or routes the request to another computational device. After execution, the output of the request is returned to the user.
\subsection{Objective Function and Definitions} 
We formulate the problem of joint service placement and task scheduling as an optimization problem, where the objective is to minimize the total cost of service provider, while satisfying all users QoS requirements. The objective function in our formulation consists of two main terms: first term represents the total cost of placing services on the cloudlets, and is comprised of our decision variable to place service $m$ on cloudlet $j$, $X_{mj}$, and the cost of making that placement: $P_{mj}^{p}$. The second term represents total cost of scheduling all tasks requested by users considering the decision variable to schedule task $t$ on cloudlet $j$, $Y_{tj}$ and the cost of placing a service of type $M(t)$ on cloudlet $j$ or cloud $C$: $P_{M(t)j}^{s}$. Considering the proposed objective function, the QoS constraint of users, and constraints on cloudlets processing and storage capacity, we formulate our problem as below. \textbf{Table 1} contains information on all relevant symbols.\\
\newcolumntype{L}[1]{>{\raggedright\let\newline\\\arraybackslash\hspace{0pt}}m{#1}}
\begin{table}
  \captionof{table}{Definitions}
  \begin{tabular} { |c|L{6.75cm}| }
  \hline
  \textbf{var} & \textbf{definition} \\ \hline
  $M$ & set of all services \\ \hline
  $M(t)$ & the service type of task $t$ \\ \hline
  $X_{mj}$ & the decision to place service $m$ on cloudlet $j$ \\ \hline
  $P^{p}_{mj}$ & cost (price) of placing service $m$ on cloudlet $j$ \\ \hline
  $T$ & set of all tasks T \\ \hline
  $t$ & task \\ \hline
  $J$ & the set of all cloudlets (including the cloud) \\ \hline
  $C$ & the cloud \\ \hline
  $J(t)$ & the cloudlet $j$ local to task $t$ \\ \hline
  $Y_{tj}$ & the decision to schedule task $t$ on cloudlet $j$ \\ \hline
  $P^{s}_{mj}$ & cost (price) of serving a task $m$ on cloudlet $j$ \\ \hline
  $P^{s}_{mC}$ & cost (price) of serving a task $m$ on the cloud  \\ \hline
  $Q_{t}$ & quality of service (time delay) threshold for task $t$ \\ \hline
  $\delta(t,j)$ & completion time of task $t$ on cloudlet $j$ \\ \hline
  $H_{m}$ & storage demands of service $m$ \\ \hline
  $S_{j}$ & total storage of cloudlet $j$ \\ \hline
  $\sigma(t)$ & computation time for task $t$ \\ \hline
  $W_{j}$ & total processing units of cloudlet $j$ \\ \hline
  $d_{jj'}$ & distance between cloudlets $j$ and $j'$ (measured in ping time)\\ \hline
  $t_{In}$ & input packet size of task $t$ \\ \hline
  $t_{Out}$ & output packet size of task $t$ \\ \hline
  \end{tabular}
  \end{table}
  
min $\sum_{m \in M}\sum_{j \in J}P^{p}_{mj}X_{mj} + \sum_{t \in T}\sum_{j \in J} P^{s}_{M(t)j}Y_{tj}$\\
s.t.
\begin{enumerate}
\item $\sum_{m \in M} H_{m}X_{mj} \leq S_{j}~~~ \forall j \in J$
\item $\sum_{t \in T} \sigma(t)Y_{tj} \leq W_{j}~~~ \forall j \in J$
\item $\sum_{j \in J} Y_{tj} = 1~~~ \forall t \in T$
\item $Y_{tj}\delta(t,j) \leq Q_{t} \forall t \in T, j \in J$
\item $\delta(t,j) = d_{J(t)j} * t_{In} + \sigma(t) + d_{J(t)j} * t_{Out}$
\item $X_{mj}, Y_{tj} \in \{0,1\}$
\item $Y_{tj} \leq X_{M(t)j}$
\end{enumerate}
\setlength\parindent{24pt}

Equations 1-2 ensure that no computational device is over capacity in storage, or overworked in terms of computation. Equation 3 ensures that each task is completed, and completed in only one location. Equation 4 dictates that each task must be completed in time to meet that task's QoS threshold. and Equation 5 details how the completion time for a user's task is calculated. Equation 6 establishes $X_{mj}$ and $Y_{tj}$ as decision variables. Finally, Equation 7 ensures that a task may only be scheduled on a computational device if that task's service is placed on that device.

\section{Methodology}
\setlength\parindent{24pt}
\subsection{Approach}
With the problem formulated as an integer linear programming problem, the intuitive next step is to outline the methodology used to solve this problem. As the goal of this research is to provide heuristics to the optimal result, it is necessary to prove that solving for the optimal solution is not feasible. First, we demonstrate that the problem is NP-hard via  a reduction to the general assignment problem. Then, we provide a brief defense of the QoS-Aware System's feasibility. Afterwards, we provide heuristics to showcase different approaches to placing services and scheduling user requests. Finally, we give a discussion of simulation parameters that are to be used check for algorithm correctness and compare performance across heuristics and the optimal solution.

\subsection{NP-Hardness}
This problem is a generalization of the general assignment problem. First, we account for all the conditions that must be met for a problem to be the general assignment problem. If we consider all placement costs to be zero, then we are tasked with assigning a number of tasks to a number of cloudlets (agents). The constraints of the general assignment problem apply to this formulation. Each agent has a constraint on how many tasks they can serve: Eq 2. Each task must be assigned exactly once: Eq 3. Lastly, the decision variable of assigning a task to an agent is binary: Eq 6. With each of the constraints of the general assignment problem accounted for, we now show that the additional constraints do not necessarily constrain the solution space of the general assignment problem. \\

Eq 1 is non-conflicting, as the storage $S_j$ can be set to $\inf$, ensuring that the condition is always met for a solution of the general assignment problem. Eq 5 is also non-conflicting, as the $Q_t$ can be set to $\inf$ much in the same way Eq 1 was handled. Lastly, Eq 7 is non-conflicting as due to our relaxation of Eq 1, $X_{M(t)j}$ can always be one (since there's no limit to the number of services we can place).

\subsection{Defense of QoS-Aware System}
To justify the inclusion of QoS-Aware systems, we show that the QoS-less system (a system not concerned with meeting the users' QoS) is not sufficient when the users have a realistic QoS-Factor. This is done by increasing the number of users in an otherwise static system and then comparing the percentage of tasks that do not meet the QoS deadline in both systems. We show the corresponding objective values to demonstrate that the inclusion of the QoS-Aware component does not grossly inflate the cost to the service provider. The values used in this simulation are outlined in \textbf{Table II}, with the exception of the number of users which is varied from fifty to five-hundred. 
\begin{figure}
  \centering
  \includegraphics[scale=.4]{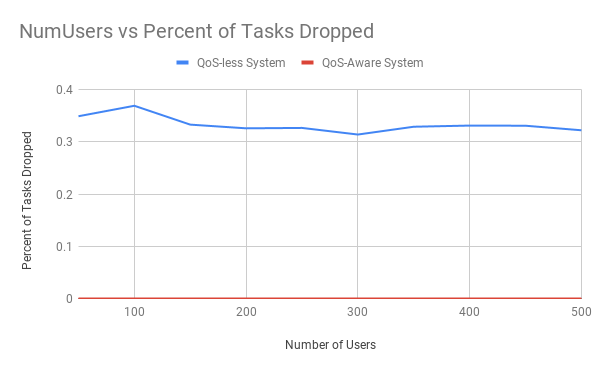}
  \caption{Number of Users vs Percentage of Tasks Dropped: Comparing QoS-Aware to QoS-less systems.}
  \label{fig:my_label}
\end{figure}
\textbf{Fig 2} shows that the QoS-less system consistently fails to complete over thirty percent of tasks, regardless of the number of users. However, this wouldn't be relevant if the inclusion of the QoS-Aware component would inflate the objective value. Thus we compare the objective values of both systems from the previous simulation and plot the results in \textbf{Fig 3}.
\begin{figure}
  \centering
  \includegraphics[scale=.4]{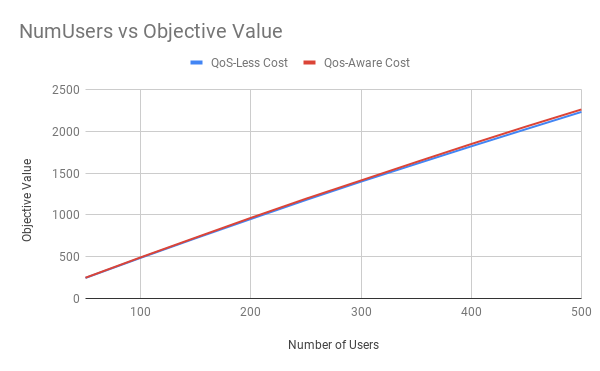}
  \caption{Number of Users vs Objective Value: Comparing QoS-Aware optimal to QoS-less optimal objective values.}
  \label{fig:my_label}
\end{figure}
In \textbf{Fig 3}, we see that the inclusion of a QoS-Aware component did not inflate the objective significantly, even at a large number of users. As the decreased percentage of tasks dropped greatly outweighs the slightly increased objective value, we conclude that the QoS-Aware system has merit worth exploring.

\subsection{Heuristics}
We consider two heuristics to approximate the optimal solution. The heuristics consider a difference of approaches including global placement/scheduling vs cloudlet-level (or local) placement/scheduling.
\begin{itemize}
  \item \textbf{Algorithm 1 Local Serving}: From requested tasks connected to that cloudlet, weigh services for that cloudlet based on the number of tasks needed that can actually be scheduled there, based on that cloudlet's constraints. Then, choose which service to place based on which task has the tightest QoS constraint. After choosing one service this way, schedule all tasks that can be scheduled. Repeat until the cloudlet is depleted of either storage or computational resources. Repeat this process for all cloudlets. Any tasks not handled by a user's local cloudlet are sent to the cloud. Line 3 begins the loop to loop over all cloudlets. Lines 4-8 detail the loop until a cloudlet has expended one of its resources. Lines 10-12 loop over all remaining tasks, scheduling them on the cloud.
  \begin{algorithm}
\caption{Local Serving}
\begin{algorithmic}[1]
\Procedure{Select and Schedule-L}{$Cl: \textrm{set of cloudlets},T: \textrm{set of tasks}$} 
  \State $ans=[]$
  \For{each $cl \in Cl$}
    \While{cl resources not expended}
      \State $t'=\{argmin(Q(t'))|conn(t',cl)\}$
      \State $ans.append([service(t'),cl])$
      \State $ans.append([t',cl])$
      \State reduce cl capacity
    \EndWhile
  \EndFor
  \State $c = cloud$
  \For{each $t \in T$ not scheduled}
    \State $ans.append([service(t),c])$
    \State $ans.append([t,c])$
  \EndFor
  \Return $ans$
\EndProcedure
\end{algorithmic}
\end{algorithm}
  \item \textbf{Algorithm 2 Global Serving}: Consider services for all computational devices based on the number of tasks needing that service. Weigh services on the number of serviceable tasks of that type. Do this for each computational device. Divide the number of serviceable tasks by the placement cost for the service on that device to find the ``profit" for that service. Choose the service/device pair with the highest profit. After choosing one service this way, schedule all serviceable tasks. Repeat this until all tasks have been scheduled. Line 3 begins the for loop to lop over all tasks. Lines 5-10 loop over each of the computational devices. Lines 7-9 loop over each service, determining how many tasks of that service can be served on the current cloudlet (accounting for hardware and time limitations), and saving that number as profit. Lines 11-12 choose which cloudlet will net the most profit, and append that to the answer. Finally, lines 13-14 schedule the answer and reduce the chosen cloudlet's capacity.
  \begin{algorithm}
\caption{Global Serving}
\begin{algorithmic}[1]
\Procedure{Select and Schedule-G}{$Cl:\textrm{set of cloudlets},T: \textrm{set of tasks}$}
  \State $ans=[]$
  \While{$\exists t \in T$ not scheduled}
    \State $s = service(t)$
    \For{each $cl \in Cl$}
      \State $profit = 0$
      \For{each t' with service s}
        \If{t' servible by cl}
          \State increase profit
        \EndIf
      \EndFor
      \State save profit and all t' servible
    \EndFor
    \State $chosen = argmax(profit)$
    \State $ans.append([s,chosen])$
    \For{all $t'$}
      \State $ans.append([t',chosen])$
        \State reduce chosen's capacities
    \EndFor
  \EndWhile
  \Return $ans$
\EndProcedure
\end{algorithmic}
\end{algorithm}
\end{itemize}

\subsection{Simulation Parameters}
\setlength\parindent{24pt}
  
\newcolumntype{M}[1]{>{\raggedright\let\newline\\\arraybackslash\hspace{0pt}}m{#1}}
\begin{table}
  \captionof{table}{Used Values}
  \begin{tabular} { |c|L{3cm}| }
  \hline
  var & value \\ \hline
  $|T|$ & 400 \\ \hline
  $|M|$ & 1000 \\ \hline
  $\beta$ & 3 \\ \hline
  $|J|$ & 4 \\ \hline
  $t_{In},t_{Out}$ & $\in \{2,4\}$ \\ \hline
  $\sigma(t)$ & $\in \{2,4\}$ \\ \hline
  $Q_{t}$ & 2.5* $\sigma(t)$\\ \hline
  $P^{s}_{mC}$ & $\in \{2,4\}$ \\ \hline
  $P^{s}_{mj}$ & $\in \{P^{s}_{mC}, \beta*P^{s}_{mC}\} s.t. j \neq C$  \\ \hline
  $P^{p}_{mj}$ & $\in \{2,4\}$ \\ \hline
  grid size & $100$ \\ \hline
  cloud distance & $5*$max grid distance \\ \hline
  \end{tabular}
  \end{table}
\setlength\parindent{24pt}
Each of the cloudlets was placed randomly on the grid and then the distances we calculated based on their geodesic location. The distance to the cloud was considered to be a multiple of the max distance of this grid. To model the preference of the service provider to send requests to the cloud, we consider the cost of scheduling a task on a cloudlet to be up to $\beta$ times the cost of scheduling on the cloud. The QoS threshold, $Q_t$, is defined as the QoS Factor times the computational time $\sigma(t)$. Given a static range of cloudlet parameters, service types, and service parameters we varied the number of users, $\beta$ value, and QoS Factor to view the changes in objective function. For a refresher, in \textbf{Table I}, is the definition of each symbol used. In \textbf{Table II}, we outline the value of each parameter for our simulations. For each number of requests, twenty different scenarios were generated in order to get a better look at the average performance as opposed to having one data point for each parameter set. All scenarios were generated using original Python scripts, solved using the Gurobi optimizer, as well as in-house software to simulate the proposed heuristics.

\section{Results} To evaluate the performance of our proposed approach, we investigate the impact of different parameters on the total cost of the system, and compared our proposed heuristics performance with the optimal solution. In the following, we showcase some of the results highlighting the impacts of number of users, the Qos constraint, and the service placement costs.   
\setlength\parindent{24pt}
\begin{figure}
  \centering
  \includegraphics[scale=.4]{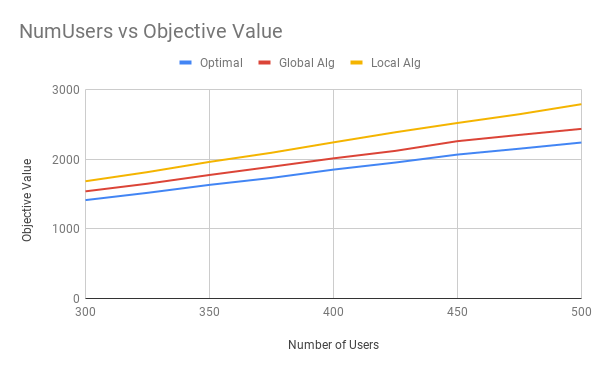}
  \caption{Number of Users vs Objective Value: Comparing optimal to local and global placement and scheduling algorithms.}
  \label{fig:my_label}
\end{figure}
\subsection{Impact of number of users}
In \textbf{Fig 4}, we increase the number of users from 300 to 500, in increments of 25. All three of the solutions show an increase in objective value as the number of users is increased, which is intuitive as each task scheduled accrues some cost to the service provider, and with more users there are clearly more tasks being scheduled.
\begin{figure}
  \centering
  \includegraphics[scale=.4]{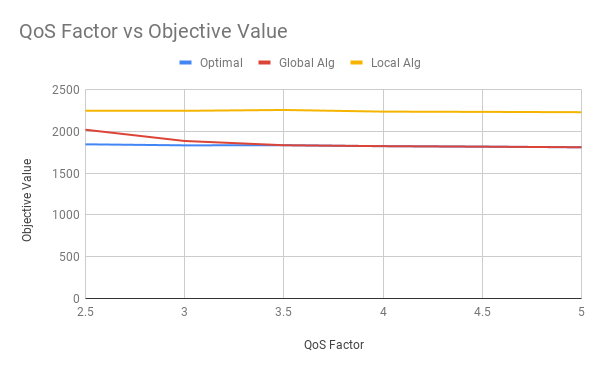}
  \caption{QoS factor vs Objective Value: Comparing optimal to local and global placement and scheduling algorithms.}
  \label{fig:my_label}
\end{figure}
\subsection{Impact of the QoS constraint}
In \textbf{Fig 5}, we increase the QoS Factor from 2.5 to 5, in increments of .5. We see, unsurprisingly, that as the QoS constraint becomes more lax, the global algorithm approaches the optimal value. This is because it is optimal for the service provider to send tasks to the cloud. As the QoS factor becomes greater, more tasks can be sent to the cloud which both the optimal and global algorithms will do. The local algorithm, conversely, will always pack the cloudlets to capacity first, meaning it cannot benefit from this relaxed constraint.
\begin{figure}
  \centering
  \includegraphics[scale=.4]{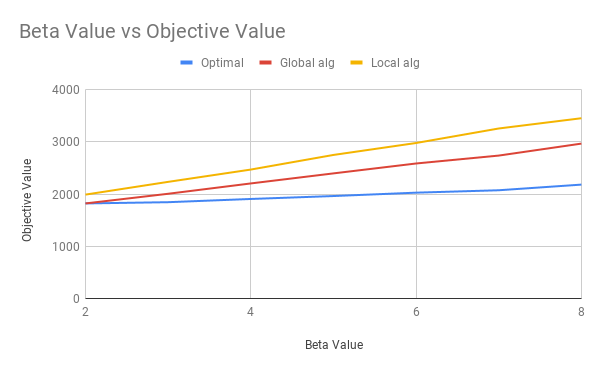}
  \caption{$\beta$ Value vs Objective Value: Comparing optimal to local and global placement and scheduling algorithms.}
  \label{fig:my_label}
\end{figure}
\subsection{Impact of scheduling cost on the edge cloud versus cloud}
\textbf{Fig 6} shows the effects that increasing the $\beta$ value has on the objective value. Recall that $\beta$ represents the price of scheduling a request on to a cloudlet (edge cloud) rather than the cloud, and it can represent the cases where a service provider has to lease a cloudlet and therefore, have to pay for using that cloudlet. We see that across all solutions, objective value increases with $\beta$ value. However, the local and global algorithm show noticeable increase in objective value. This is because of their increased likelihood to schedule tasks on the cloudlets, especially in the case of the local algorithm. This is backed up by the local value obtaining the largest objective value, as it schedules the most tasks on the cloudlets of the three algorithms.

Across all three simulations, we see the same relative behavior of the solutions. In all cases, the global algorithm outperforms its local counterpart. This demonstrates the need for cloudlets to be cognizant of the demands on the surrounding cloudlets if cost to the service provider is to minimized. This finding resonates with the intuition of scheduling to allow cloudlets to serve not only their local requests, but remote requests which they are already configured to serve.

\section{Conclusion}
Over the course of this paper, we introduced the need for a QoS-Aware system in a multi-tiered cloud, and win-win solutions for users and service providers. We examined problems similar to this in the existing literature, noting edge-computing and service placement as particular areas of interests, and asserted the novelty of our configuration. The problem was then formulated as an ILP, and proven to be NP-Hard. We defend the assertion of QoS-Aware need by showing that with a sufficient QoS factor, the QoS-Aware solution is not significantly worse than the QoS-less solution in terms of the provider cost, while the QoS-less solution has to drop some of the user's request and the QoS-Aware solution does not drop any user requests. Heuristics were introduced, one which considered each computational device independently, and one which considered computational devices as pieces of a global system. Our simulations then showed that the global approach outperforms the local approach in terms of cost to the service provider, and in some situations is very near optimal.

In the future, we would like to consider a more realistic request schedule where requests are entered at dynamically and the system adapts to requests as they are received. We would also like to consider perturbation in the estimated calculation time, and geographically dynamic users.

\bibliographystyle{plain}
\bibliography{refs.bib}
\end{document}